\documentclass[english,12pt]{article}
\usepackage{color}
\usepackage{amssymb}
\usepackage{amsmath}
\usepackage{babel}
\usepackage{pifont}
\usepackage
[dvips,
colorlinks=true,
linkcolor=webgreen,%defined below
filecolor=webbrown,%defined below
citecolor=webgreen,%defined below
bookmarksopen=false,
]{hyperref}
\definecolor{webgreen}{rgb}{0,.5,0}
\definecolor{webbrown}{rgb}{.6,0,0}
\usepackage[T1]{fontenc}
\usepackage[cp1250]{inputenc}
\usepackage{array}
\usepackage[dvips]{graphicx}
\usepackage{amssymb}
\date{}
\definecolor{arcolor}{cmyk}{0.05,0.95,0.9,0.1}
\title{Giffen paradoxes in quantum market games}
\author{Jan S\l adkowski \\ Institute of Physics, University of Silesia, \\ Uniwersytecka
4, Pl 40007 Katowice, Poland \\ e-mail:
\href{mailto:sladk@us.edu.pl}{sladk@us.edu.pl} }
\begin{document}
\baselineskip8mm \maketitle
\def\Z{{\bf Z\!\!Z}}
\def\R{{\bf I\!R}}
\def\N{{\bf I\!N}}
\def\C{{\bf I\!\!\!\! C}}
\begin{abstract}
\baselineskip7mm Recent development in quantum computation and
quantum information theory  allows to extend the scope of game
theory for the quantum world. The paper presents the history and
basic ideas of quantum game theory. Description of Giffen
paradoxes in this new formalism  is discussed.

\end{abstract}


 \vspace{5mm}

\section{Motivation}
Attention to the very physical aspects of information
characterizes  the recent research in quantum computation, quantum
cryptography and quantum communication.  In most of the analysed
cases quantum description of the system provides advantages over
the classical situation. The flagships of quantum information are:
famous  Shor's polynomial time quantum algorithm for factoring
\cite{1}, Simon's quantum algorithm to identify the period of a
function chosen by an oracle (more efficient than any
deterministic or probabilistic algorithm) \cite{2} and the quantum
protocols for key distribution, devised by Wiener, Bennett and
Brassard, and Ekert (qualitatively more secure against
eavesdropping than any classical cryptographic system) \cite{3,4}. \\

Game theory, the study of (rational) decision making in conflict
situation, seems to ask for a quantum version. Games against
nature, originally studied by Milnor \cite{6}, include those for
which nature is quantum mechanical. Many of quantum information
problems have game-theoretic counterparts.  Finally, von~Neumann
is one of the founders of both modern game theory \cite{7} and
quantum theory. Classical strategies can be pure or mixed: why
cannot they be entangled or interfere with each other? Can quantum
strategies  be more successful than classical ones? Are they of
any practical value?

\section{Quantum Games}
Any quantum system which can be manipulated by two  or more
parties, and where some utility of the moves can be reasonably
defined, may be conceived as a quantum game \cite{8}-\cite{10}.
For example, a {\it two-player quantum game}\/ $\Gamma=({\cal
H},\rho,P_A,P_B)$ is completely specified by the underlying
Hilbert space ${\cal H}$ of the physical system, the initial state
$\rho\in {\cal S}({\cal H})$, where ${\cal S}({\cal H})$ is the
associated state space and $\rho=\rho_A\otimes\rho_B$  describes
the players, say Alice (A) and Bob (B), initial strategies
$\rho_A$ and $\rho_B$. The { pay-off (utility) functions}\/ $P_A$
and $P_B$ specify the pay-off for each player. {\it Quantum
tactics}\/  $S_A$ and $ S_B$  are linear (quantum) operations,
that is, a completely positive trace-preserving map mapping the
state space on itself. Employing a tactics, that is performing the
appropriate linear map, describes a change of the players
strategy. The quantum game's definition may also include certain
additional rules, such as the order of the implementation of the
respective quantum strategies. We also exclude the alteration of
the pay-off during the game. The generalization for the N players
case is obvious. Schematically we have:
$$\rho\stackrel{(S_A,S_B)}{\longmapsto}\sigma\Rightarrow(P_A,P_B).$$

 \section{Quantum Market Games }

It is tempting to check if quantum game theory may  be suitable
for description of market transactions. A quantum game like
description of market phenomena in terms of supply and demand
curves was proposed in Ref. \cite{11}-\cite{13}. In this approach
quantum strategies are vectors in some Hilbert space and can be
interpreted as superpositions of trading decisions. For an
economist (or trader) they form the potential "quantum board".
Strategies and not the apparatus nor the installation for actual
playing  are at the very core of the theory. If necessary the
actual subject of investigation may consist of single traders,
teams of traders or even the whole market. Due to the possible
economics context the quantum strategies reveal a lot of
interesting properties. Supply strategies of market objects are
Fourier transforms of their respective demand states \cite{13}. \\

 Of course, sophisticated equipment built according to quantum
rules may be necessary for generating or clearing quantum market
but we must not exclude the possibility that human consciousness
(brain) performs that task equally well. Even more, a sort of
quantum playing board may be the natural theater of "conflict
games" played by our consciousness. The agents (market players)
strategies are described in terms state vectors $|\psi\rangle$
belonging to some Hilbert space ${\cal H}$ \cite{10,12}. The
probability densities of revealing the agents, say Alice and Bob,
intentions are described in terms of random variables $p$ and $q$:
\begin{equation}
\label{eigenstosc} \frac{|\langle q|\psi\rangle_A|^2}{\phantom{}_A
\langle\psi|\psi\rangle_A}\, \frac{|\langle
p|\psi\rangle_B|^2}{\phantom{}_B \langle\psi|\psi\rangle_B}\;d q d
p\, ,
\end{equation}
where $\langle q|\psi\rangle_A$\/ is the probability amplitude of
offering the price $q$ by Alice who wants to buy and the demand
component of her state is given by
$|\psi\rangle_A\in\mathcal{H}_{A}$\/. Bob's amplitude $\langle
p|\psi\rangle_B$\/ is interpreted in an analogous way (opposite
position). A short look at error theory (second moments of a
random variable describe errors), Markowitz's portfolio theory and
L. Bachelier's theory of options (the random variable $q^{2} +
p^{2}$ measures joint risk for a stock buying-selling transaction)
suggest the following definition of   {\it the risk inclination
operator} (a quantum observable):
$$
H(\mathcal{P}_k,\mathcal{Q}_k):=\frac{(\mathcal{P}_k-p_{k0})^2}{2\,m}+
                     \frac{m\,\omega^2(\mathcal{Q}_k-q_{k0})^2}{2}\,,
\eqno(2) $$ \noindent where $p_{k0}:=\frac{
\phantom{}_k\negthinspace\langle\psi|\mathcal{P}_k|\psi\rangle_k }
{\phantom{}_k\negthinspace\langle\psi|\psi\rangle_k}\,$,
$q_{k0}:=\frac{
\phantom{}_k\negthinspace\langle\psi|\mathcal{Q}_k|\psi\rangle_k }
{\phantom{}_k\negthinspace\langle\psi|\psi\rangle_k}\,$,
$\omega:=\frac{2\pi}{\theta}\,$.  $ \theta$ denotes the
characteristic time of transaction \cite{12} which is, roughly
speaking, an average time spread between two opposite moves of a
player (e.~g.~buying and selling the same asset). The parameter
$m>0$ measures the risk asymmetry between buying and selling
positions.\\

Analogies with quantum harmonic oscillator allow for the following
characterization of quantum market games. The constant $h_E$
describes the minimal inclination of the player to risk. It is
equal to the product of the lowest eigenvalue of
$H(\mathcal{P}_k,\mathcal{Q}_k) $ and $2\theta$. $2\theta $ is in
fact the minimal interval during which it makes sense to
measure the profit \cite{11}.\\

Except the ground state all the strategies
$H(\mathcal{P}_k,\mathcal{Q}_k)|\psi\rangle={const}|\psi\rangle$
are {\it giffens}  that is goods that do not obey the law of
demand and supply, see bellow.  It should be noted here that in a
general case the operators $\mathcal{Q}_k $ do not commute because
traders observe moves of other players and often act accordingly.
One big bid can influence the market at least in a limited time
spread. Therefore it is natural to apply the formalism of
noncommutative quantum mechanics where one considers
$$ [ x^{k},x^{l}] = i \Theta ^{kl}:=i\Theta \,\epsilon ^{kl}.\eqno(3) $$
The analysis of harmonic oscillator in more then one dimensions
 imply that the parameter $\Theta $ modifies the constant
$\hslash_E$ $\rightarrow \sqrt{\hslash_E^{2} + \Theta ^{2}} $ and,
accordingly, the eigenvalues of $H(\mathcal{P}_k,\mathcal{Q}_k)$.
This has the natural interpretation that moves performed by other
players can diminish or increase one's inclination to taking risk.
\section{Market as a measuring apparatus } When a game allows a
great number of players in then it is useful to consider it as a
two-players game: the trader $|\psi\rangle_{k}$ whom we are
observing against the Rest of the World (RW). The concrete
algorithm $\mathcal{A}$ that is used for clearing the market may
allow for an effective strategy of RW (for a sufficiently large
number of players the single player strategy should not influence
the form of the RW strategy). If one considers the RW strategy it
make sense to declare its simultaneous demand and supply states
because for one player RW is
a buyer and for another it is a seller.\\

To describe such situations it is convenient to use the Wigner
formalism. The pseudo-probability $W(p,q)dpdq$ on the phase space
$\{(p,q)\}$ known as the Wigner function is given by
\begin{eqnarray*}
W(p,q)&:=& h^{-1}_E\int_{-\infty}^{\infty}e^{i\hslash_E^{-1}p x}
\;\frac{\langle
q+\frac{x}{2}|\psi\rangle\langle\psi|q-\frac{x}{2}\rangle}
{\langle\psi|\psi\rangle}\; dx\\
&=& h^{-2}_E\int_{-\infty}^{\infty}e^{i\hslash_E^{-1}q x}\;
\frac{\langle
p+\frac{x}{2}|\psi\rangle\langle\psi|p-\frac{x}{2}\rangle}
{\langle\psi|\psi\rangle}\; dx,
\end{eqnarray*}
where the positive constant $h_E=2\pi\hslash_E$ is the
dimensionless economic counterpart of the Planck constant. Recall
that this measure is not positive definite except for very special
cases. In a general case the pseudo-probability density of RW is a
countable linear combination of appropriate Wigner functions,
$\rho(p,q)=\sum_n w_n W_n (p,q)$,
 $w_n\geq 0$, $\sum_n w_n =1$.
The diagrams of the integrals of the RW pseudo-probabilities
$$
F_d(\ln c):=\int_{-\infty}^{\ln c} \rho(p={const.},q)dq \eqno(4)
$$ (RW bids selling at $\exp {(-p)}$)\\and
$$
F_s(\ln c):=\int_{-\infty}^{\ln \frac{1}{c}}
\rho(p,q={const.})dp\eqno(5)
$$ (
RW bids buying at $\exp{q}$ ) against the argument $\ln c$ may be
interpreted as the dominant supply and demand curves in Cournot
(French) convention, respectively \cite{13}. Note, that due to the
lack of positive definiteness of $\rho $, $F_d$ and $F_s$ may not
be monotonic functions. Textbooks on economics give examples of
such departures from the low of supply  and demand (Giffen
paradox). Fashion business and work supply are the source of everyday examples of such assets. \\

%For future reference we propose to call an arbitrage algorithm
%resulting in non positive definite probability densities {\it a
%giffen} and a strategy resulting in such market behaviour {\it a
%giffen strategy}.

\section{Giffen paradoxes}
\begin{figure}[h]
\begin{center}
\includegraphics[height=8.25cm, width=12cm]{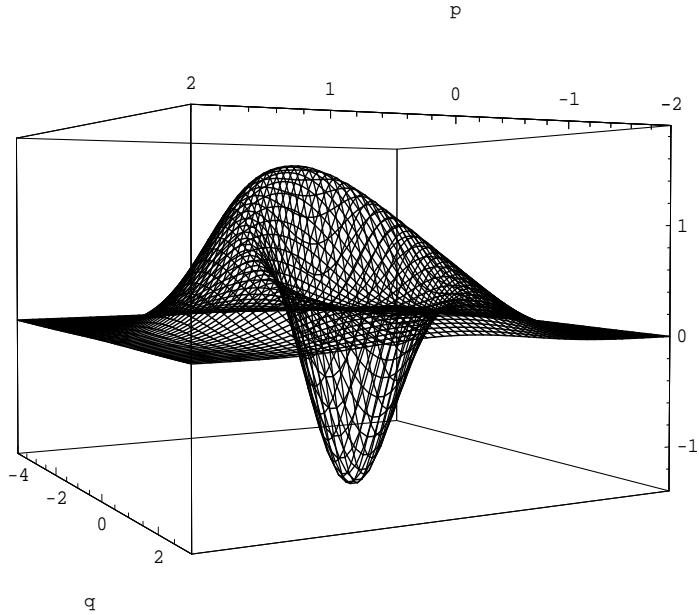}
\end{center}
\caption{ Exemplary plot of a Wigner function .} \label{interpiec}
\end{figure}

\begin{figure}[h]
\begin{center}
\includegraphics[height=6.25cm, width=9cm]{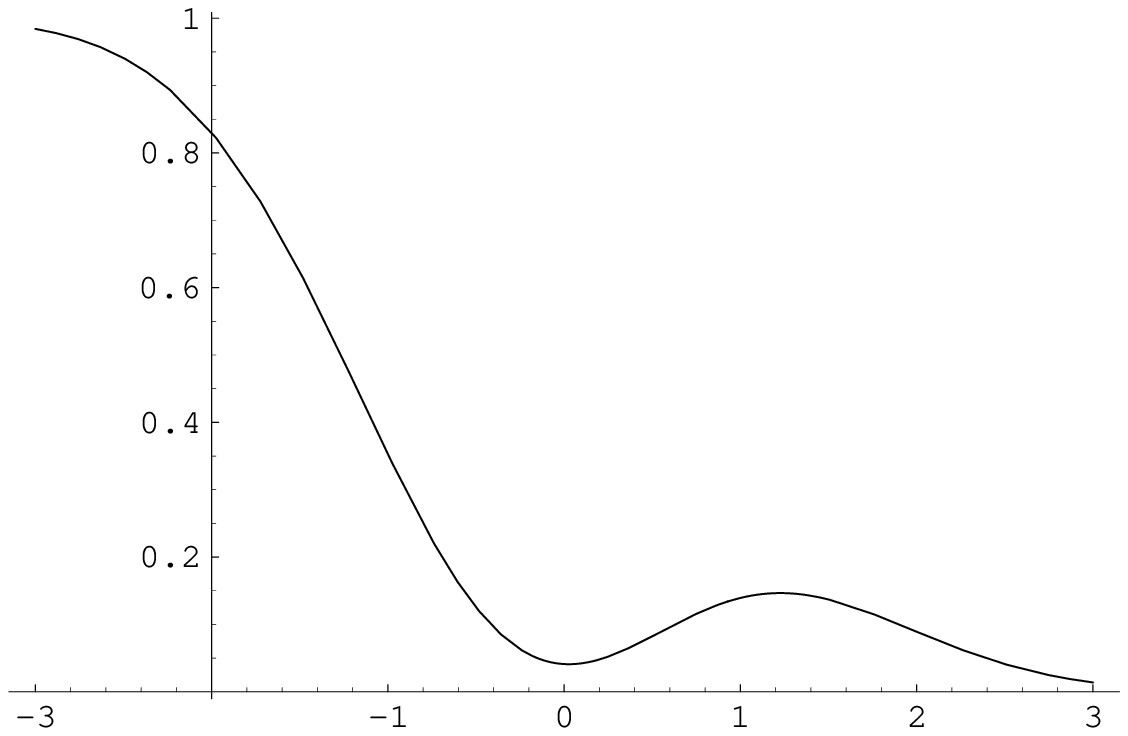}
\end{center}
\caption{Non-monotonous  conditional demand: the integral curve
for the intersection of the surface from Figure~\ref{interpiec}
with the plain $p=0.4[\frac{\hslash_E}{\sigma}]$).}
\label{interszes}
\end{figure}
 Note that the asymmetric crater-like hollow in
(Figure~\ref{interpiec}) has the minimum bellow zero, the fact
which qualitatively distinguishes the Wigner function from the
supply and demand distributions for models formulated in the realm
of the classical probability theory in which the measure of the
probability has to
be nonnegative. \\

The intersection of the surface of the diagram with the surface
given by $p\negthinspace=\negthinspace constant $ represents the
conditional probability density which is the measure of the
probability for the withdrawal price of the player in the
situations when this price is constant during the act of selling.
The withdrawal price is defined as the maximal (minimal) price the
player is going to pay (obtain) for the asset in question
\cite{14,15,13}. \\

The cross sections for the negative values of the Wigner function
are characteristic for the situation of a giffen strategy. The
suitable integrals for these curves represent fully rational
situations for which the demand (or supply) cease to be a
monotonous  function. The example of such a reaction of the player
(it might be the rest of the world) is illustrated in
Figure~\ref{interszes}\index{The rest of the world @{\em reszta
świata}\/ (RŚ)}.  We observe here the lack of the property of the
monotonicity for the demand (or supply) curves (Giffen paradox).
%turns out here to
%be the consequence of the desire to answer the question connected
%with the conditional probability for everyone of the non-Gausian
%strategy only.
In this context it is worthy to raise the question whether the
legendary captain Giffen, after observing a market anomaly which
is contradictory to the law of demand, has recorded the surprising
(although having  logical explanation) demand that decreases after
the fall of the price, or simply noticed
%the alteration
%between two of the standard\footnote{Which are in agreement with
%the law of demand} demand curves only, which took place on the
%market under
the destructive interference which had been the effect of a
careful demand transformation characteristic for a intelligent
(hence acting rationally) but poor consumer \cite{16}. The authors
incline towards the second answer. It has the advantage of being
capable of  falsification  which is a consequence of the
 precision   qualitative predictions for this
phenomenon made by the quantum theory. \\

Therefore it seems important to look after the conditions of the
market  under which the strategies described by  normal
distributions do not lead to the maximization of value of the
intensity of the gain \footnote{see also E. W. Piotrowski's
lecture in current issue} \cite{14}. They might explain the
circumstances in which we met the Giffen paradoxes.
\section{Summary and outlook}

 All this tempt us into formulating the {\em quantum
anthropic principle} of the following form.  At earlier
civilization stages markets are governed by classical laws (as
classical logic prevailed in reasoning) but the incomparable
efficacy of quantum algorithms in multiplying profits will result
in continuous change in human attitude towards quantum information
processing.  The growing significance of quantum phenomena in
modern technologies and their influence on economics will result
in quantum behaviour prevailing over the classical one. Therefore
we envisage markets cleared by quantum algorithms (computers),
quantum auctions providing agents with new means \cite{15} and
quantum games being important tools in social sciences, economics
and biology \cite{15}-\cite{21}.
%\begin{itemize}
%\end{itemize}

%\vspace*{1cm}

\end{document}